\documentclass[cameraready]{Interspeech}
\usepackage{makecell}
\usepackage{algorithm}
\usepackage{algpseudocode}
\title{AQA-TTRL: Self-Adaptation in Audio Question Answering with Test-Time Reinforcement Learning}

\author[affiliation={1}]{Haoyu}{Zhang}
\author[affiliation={1}]{Jiaxian}{Guo}
\author[affiliation={1}]{Dong}{Yang}
\author[affiliation={1}]{Yusuke}{Iwasawa}
\author[affiliation={1}, correspondingauthor]{Yutaka}{Matsuo}
\address{
    $^1$ The University of Tokyo, Japan 
}

\email{\{haoyu.zhang, jiaxian.guo, iwasawa, matsuo\}@weblab.t.u-tokyo.ac.jp, ydqmkkx@gmail.com}


\usepackage{comment}

\begin{document}

\maketitle

\begin{abstract}
    Large Audio Language Models (LALMs) exhibit strong capabilities in general audio understanding but remain static after deployment, limiting their adaptability to real-world data. Since supervised fine-tuning is costly, we propose AQA-TTRL, a novel framework for audio understanding that enables on-the-fly evolution via test-time reinforcement learning using only unlabeled test data. It generates pseudo-labels via majority voting and optimizes the model through reinforcement learning. To address the noise in self-generated labels, we introduce confidence weighting to adjust training signals. Furthermore, multiple-attempt sampling mitigates advantage collapse and stabilizes training. Across MMAU, MMAR, and MMSU, AQA-TTRL achieves significant average improvements of 4.42\% for Qwen2.5-Omni 7B and 11.04\% for the 3B model. Notably, the adapted 3B model outperforms direct inference of the unadapted 7B model, highlighting the effectiveness of test-time adaptation in audio understanding.
    
\end{abstract}

\section{Introduction}
\label{sec:intro}

The application of Large Audio Language Models (LALMs) has led to notable improvements in audio understanding~\cite{gong2023listen,tang2024salmonn,ghosh2025audio_3,lu2025desta25Audio,Qwen2.5-Omni,wu2025step}.
In practical speech and audio applications, test-time inputs typically vary due to background noise, recording conditions, and speaker variability, leading to acoustic mismatch that degrades AQA performance. 
A straightforward solution is to collect and annotate new audio-question pairs for supervised updates; however, this process is costly and time-consuming, motivating the need for label-free adaptation mechanisms.

\begin{figure}[t]
    \centering
    \includegraphics[width=0.92\linewidth]{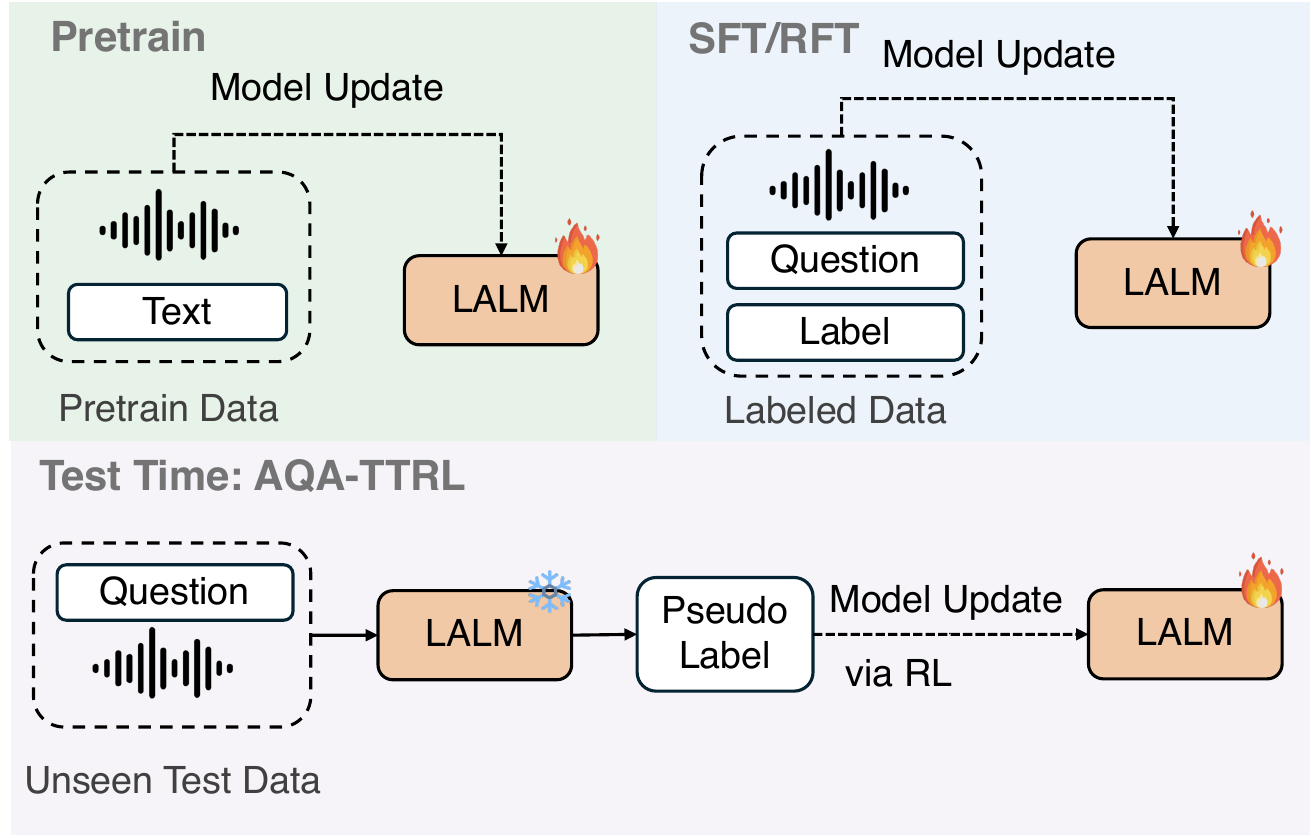}
    \caption{Role of AQA-TTRL: When faced with unseen test data, AQA-TTRL enables an automatic, label-free adaptation pipeline where model updates are driven by self-generated pseudo-labels, without any human annotation or external data.}
    \label{fig:overall} 
\end{figure}

Enabling LALMs to self-evolve on unlabeled test data at inference time provides a promising direction, mirroring how humans adapt to new situations without explicit instruction. 
While Test-Time Reinforcement Learning (TTRL)~\cite{zuo2025ttrl} has shown the potential for such self-improvement in mathematical domains, its application to the complexities of audio remains unexplored. 
In this paper, we focus on improving performance in Audio Question Answering (AQA) tasks by introducing AQA-TTRL, a novel self-adaptation method that builds a closed-loop for self-improvement (Fig.~\ref{fig:overall}). It first establishes its consensus on an answer by performing majority voting over multiple outputs, creating a pseudo-label that acts as a self-generated reward signal. This signal then guides policy optimization using Group Relative Policy Optimization (GRPO)~\cite{deepseek-math,guo2025deepseek}. To establish a reliable self-improvement loop with the inherent noisy pseudo-label, we further introduce two innovations: (1) a confidence-weighting scheme to prioritize high-quality internal signals by reweighting the training gradients, and  (2) a multiple-attempt sampling strategy to mitigate the ``advantage collapse" problem, where advantage degenerates to zero and model updates stagnate.

We provide empirical evidence that this self-improvement process is highly effective. Experimental results demonstrate that AQA-TTRL achieves an average improvement of 4.42\% on Qwen2.5-Omni 7B and 11.04\% on the 3B model across four benchmarks. Notably, the adapted 3B model surpasses the static 7B model. In summary, our contributions are:
\begin{itemize}
  \item We propose AQA-TTRL, a novel self-adaptation method for model self-improvement in audio understanding, leveraging confidence-weighted advantage and a novel sampling strategy to create a stable and effective test-time learning loop.
  \item To our knowledge, we provide the first empirical evidence that LALMs can achieve significant self-improvement in AQA tasks, demonstrating this new paradigm on the MMAU (test-mini/test)~\cite{sakshi2025mmau}, MMAR~\cite{ma2025mmar}, and MMSU~\cite{wang2025mmsu}.
  \item We justify our approach by demonstrating a strong positive correlation between model confidence and correctness on test data, validating the core mechanism that drives our self-improvement framework.
\end{itemize}

\section{Related works}
\label{sec:relatedwork}

\textbf{Large audio language models}. 
Large Audio Language Models (LALMs) demonstrate strong abilities in understanding and reasoning across a variety of tasks involving auditory input, such as speech, sound, and music.
They typically use LLMs as their backbone for understanding and text generation, while incorporating an audio encoder to enable auditory perception. Representative models include LTU~\cite{gong2023listen,gong_ltuas}, SALMONN~\cite{tang2024salmonn}, Audio Flamingo series~\cite{ghosh2025audio_3,kong2024audio,ghosh2025audio}, DeSTA series~\cite{lu2025desta25Audio,lu24c_interspeech,lu2025developing}, Step-Audio series~\cite{wu2025step,huang2025step}, Qwen Audio series~\cite{Qwen-Audio,Qwen2-Audio} and Qwen Omni series~\cite{Qwen2.5-Omni,Qwen3-Omni}. 
By formulating diverse tasks as text generation, LALMs overcome task-specific constraints.
These works improve the accuracy on traditional audio tasks and are capable of processing open-ended Audio Question Answering (AQA).

\textbf{Reinforcement learning with verifiable rewards}. 
Reinforcement Learning with Verifiable Rewards (RLVR) has been proven effective in improving LLMs. 
DeepSeek Math~\cite{deepseek-math} and DeepSeek-R1-Zero~\cite{guo2025deepseek} use RLVR to improve reasoning ability without reward models. 
SEED-GRPO~\cite{chen2025seed} applies semantic entropy to represent uncertainty for calculating the advantage.
TTRL~\cite{zuo2025ttrl} uses RLVR with majority voting at test time to boost performance on math benchmarks.
Spurious Rewards~\cite{shao2025spurious} shows that even weak or spurious labels can improve performance with RLVR. 
For the combination of LALMs and RLVR, R1-AQA~\cite{li2025reinforcement} shows that GRPO can outperform supervised finetuning on MMAU benchmarks.
Omni-R1~\cite{rouditchenko2025omni} achieves the SOTA performance on MMAU and MMAR by applying GRPO with an LLM augmented dataset.
In contrast to prior work that relies on labeled training data for RL post-training, our method is under a label-free training paradigm.

\section{Method}
\label{sec:method}
In this section, we first present the problem setting of test-time adaptation using unlabeled test data.
Next, we present an overview of our method, covering pseudo-label generation and model update, followed by its two key innovations: confidence-weighted advantage and multiple-attempt sampling. 

\subsection{Problem setting}
\label{ssec:prblem_setting}
Audio understanding models are often improved by collecting domain audios and questions, annotating them, and updating the model with labeled pairs.
However, this procedure is costly, making full automation difficult. Moreover, domain data collection may induce a distribution shift relative to real-world test data, potentially affecting robustness.
If the model can adapt itself at test time \textbf{with unlabeled test data only}, it not only reduces annotation costs but also facilitates adaptation to new test conditions without additional annotation.
In this work, we focus on Audio Question Answering (AQA) as a representative task, and our problem setting is to improve AQA performance through test-time adaptation under a label-free paradigm.

\subsection{Overview of the AQA-TTRL}
\label{ssec:pipeline}

\begin{figure}[t]
\centering
\includegraphics[width=\linewidth]{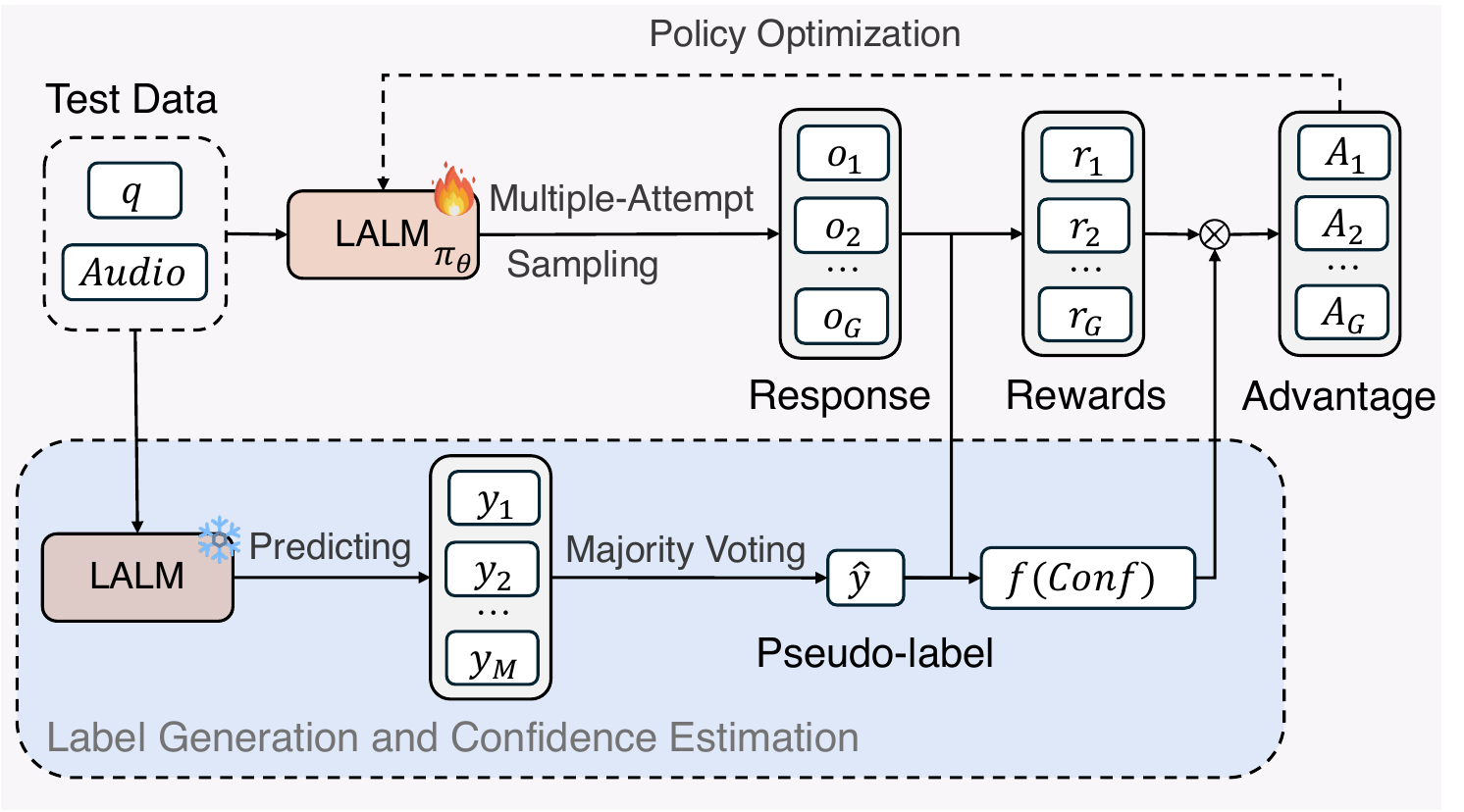}
\caption{Overview of AQA-TTRL. The framework derives pseudo-labels with confidence through majority voting, conducts multiple-attempt sampling for effective response generation, and scales the learning signal by confidence.}
\label{fig:ttrl} 
\end{figure}

In order to improve the model quality without annotated data, AQA-TTRL contains two stages, as Fig.~\ref{fig:ttrl} shows: 1) Pseudo-label generation. 2) Pseudo-label Guided Model Updates. 
% Motivated by \textit{Self-Consistency}~\cite{wang2022self}, the model first generates multiple predictions $\{y_1, y_2, \ldots, y_M\}$ for a given input, and then conducts majority voting on these responses to establish a consensus answer, which serves as the pseudo-label from the model's own consensus:
Motivated by \textit{Self-Consistency}~\cite{wang2022self}, for each audio-question pair $(a,q)$, the model first generates $M$ predictions $\mathcal{Y}=\{y_1,y_2,\ldots,y_M\}$. It then conducts majority voting over candidate answers $c \in \mathcal{Y}$ to aggregate these stochastic predictions into a consensus pseudo-label:
\begin{equation}
\hat{y} = \arg\max_{c \in \mathcal{Y}} \textstyle \sum_{m=1}^M \mathbf{1}\{ y_m = c \}, 
\quad \hat{y}: \text{pseudo-label}.
\end{equation}

For model updates, we adopt Group Relative Policy Optimization (GRPO) \cite{deepseek-math,guo2025deepseek}.
Unlike supervised fine-tuning, which directly imitates noisy pseudo labels, GRPO optimizes reward signals and provides a more noise-robust foundation for test-time adaptation.
For each audio-question pair, the policy model $\pi_\theta$ generates a set of responses $\{o_1, o_2, \ldots, o_G\}$, each referred to as a rollout. Subsequently, the reward score $r_i$ is calculated by comparing the response $o_i$ with the pre-generated pseudo-label $\hat{y}$ using the reward function. Following prior work~\cite{li2025reinforcement}, we adopt two binary exact-match rewards: a format reward and an accuracy reward,
defined as $r_i = r_{acc}(o_i,\hat{y}) + r_{format}(o_i)$

The advantage $A_{i}$ is derived by normalizing the rewards $\{r_1, r_2, \ldots, r_G\}$ within each group. The policy model $\pi_\theta$ is then optimized by maximizing the following objective:
\begin{equation}
\mathcal{J}_{\mathrm{GRPO}}(\theta) = \mathbb{E}_{q, a, \{o_i\}} \Bigg[ \frac{1}{G} \sum_{i=1}^G \frac{1}{|o_i|} \sum_{t=1}^{|o_i|} ( \mathcal{O}_{i,t} - \beta d_{i,t} ) \Bigg],
\end{equation}
\vspace{-1.5em}
\begin{align*}
&\text{where} \\[-0.4em] 
&\mathcal{O}_{i,t} = \min \bigg( c_{i,t}(\theta) A_{i}, \mathrm{clip}\Big(c_{i,t}(\theta), 1-\epsilon, 1+\epsilon\Big) A_{i} \bigg), \\[-0.1em] 
&c_{i,t}(\theta) = \frac{\pi_{\theta}(o_{i,t} \mid q, a,o_{i,<t})}{\pi_{\theta_{\mathrm{old}}}(o_{i,t}\mid q, a, o_{i,<t})}, \quad 
d_{i,t} = D_{\mathrm{KL}} \left( \pi_{\theta} \parallel \pi_{\mathrm{ref}} \right).
\end{align*}
Here $a$ denotes the audio input, and $\epsilon$, $\beta$ are hyperparameters that limit per-iteration deviation from the base model.

However, label-free test-time adaptation with pseudo-labels faces two challenges. First, the self-generated pseudo-labels are inherently noisy, and learning from them directly is not desirable. Second, high-confidence samples tend to produce identical rollouts, making rewards constant within a group and causing the normalized advantage to vanish (advantage collapse).
To address these challenges, we introduce two coordinated mechanisms that jointly restore stable and reliable learning dynamics:
(1) Confidence-weighted advantage (Sec.~\ref{ssec:confidence}), which re-scales normalized gradients according to pseudo-label reliability, aligning update magnitude with signal quality;
and (2) Multiple-attempt sampling (Sec.~\ref{ssec:MAS}), which detects rollout collapse and sequentially selects the first non-identical group, thereby mitigating advantage collapse.
\subsection{Confidence-weighted advantage}
\label{ssec:confidence}

LALMs output answers in text without explicit confidence scores. 
We estimate pseudo-label confidence via majority-vote consistency,
$Conf = \frac{1}{M}\sum_{m=1}^M \mathbf{1}\{y_m=\hat y\}$,
and use it to weight the normalized advantage:
\begin{equation}
A_{i} = \frac{r_i - mean(\{r_1,r_2,...,r_G\})}{std(\{r_1,r_2,...,r_G\})} \cdot f(Conf).
\end{equation}

We use $f(Conf)=\exp(Conf)$ in all experiments; linear and square-root were also tested, with the exponential variant performing best overall.
We build on the observation that higher prediction confidence is generally associated with higher label reliability. We therefore use confidence for relative reweighting: higher-confidence pseudo-labels receive larger weights than lower ones. Moreover, the exponential mapping yields a bounded weight range ($[1,\mathrm{e}]$), which preserves the difference across confidence levels while limiting over amplification from potentially miscalibrated high-confidence pseudo-labels.

\subsection{Multiple-attempt sampling}
\label{ssec:MAS}

\vspace{-0.5em}
\begin{algorithm}
\caption{Multiple-attempt sampling}
\label{alg:mas}
\begin{algorithmic}[1]
\State {\bf input:} policy $\pi_\theta$, old policy $\pi_{\theta_{\text{old}}}$, group size $G$, pseudo-labeled dataset $D$ 
\For{each $(q, a, \hat{y}, Conf) \in D$}
    \State Generate candidates $\mathcal{G}_1, \mathcal{G}_2, \mathcal{G}_3 \sim \pi_\theta(\cdot \mid q, a)$
    \If{\textbf{not} all\_same($\mathcal{G}_1$)}
        \State $\{o_i\}_{i=1}^G \gets \mathcal{G}_1$
    \ElsIf{\textbf{not} all\_same($\mathcal{G}_2$)}
        \State $\{o_i\}_{i=1}^G \gets \mathcal{G}_2$
    \Else
        \State $\{o_i\}_{i=1}^G \gets \mathcal{G}_3$
    \EndIf
    \State (followed by optimization steps)
\EndFor
\end{algorithmic}
\end{algorithm}

When a limited number of responses are sampled, high-confidence questions may collapse to identical outputs,
leading to vanishing advantage and stalled updates.
Although increasing the group size can reduce collapse, incorporating audio tokens in LALMs increases sequence length and GPU memory consumption, making large groups memory-prohibitive. 
We therefore adopt sequential multi-attempt sampling. 
As illustrated in Algorithm~\ref{alg:mas}, multiple response groups with equal size are generated, and the first group with non-identical samples is selected, using later groups as fallback when earlier ones collapse.
Since high-confidence samples are generally of higher quality, leveraging them is crucial for effective self-adaptation.
While this strategy does not guarantee diversity, it mitigates the issue under a limited computing budget. 
Combined with confidence weighting, it preserves informative training signals.

\section{Experiments}
\label{sec:experiment}

In this section, we introduce our experimental settings and report performance on the following AQA benchmarks: MMAU-v05.15.25 (test-mini/test), MMAR, and MMSU. 

\subsection{Experimental settings}
\label{ssec:subhead}
We use Qwen2.5-Omni 7B and 3B \cite{Qwen2.5-Omni} as base models for test-time adaptation. The pseudo-label is obtained via majority voting over 64 predictions (T=1) in advance. For RL-based adaptation, we perform GRPO rollouts with 4 generations (T=1) and run 500 update steps for all datasets. 
We report results at step 100 for smaller datasets (MMAU test-mini and MMAR), and at step 500 for larger datasets (MMAU test and MMSU). 
We fine-tune all model parameters using AdamW (lr=$1\mathrm{e}{-6}$, weight decay=0.01) with a global batch size of 8 (4 GPUs, micro-batch=1, gradient accumulation=2) and gradient clipping of 1.0 in bf16. Following prior GRPO practice~\cite{shao2025spurious,zhang2026scafgrpo}, the $\epsilon$ is 0.2 and $\beta$ is 0. The prompt is: ``\textit{\{question\} Please choose the answer from the following options: \{choice string\}. Output the final answer in $<$answer$>$ $<$/answer$>$.}'' Similar to the previous research~\cite{li2025reinforcement, rouditchenko2025omni}, we disable thinking and let the model output the answer directly. We additionally include a Supervised Fine-Tuning (SFT) baseline that optimizes cross-entropy on the same majority-vote pseudo-labels for 3 epochs.

\subsection{Main results across different benchmarks}
\label{ssec:result}

\vspace{-0.5em}
\begin{table}[th]
\centering
\caption{Accuracies across different benchmarks: DI = Direct Inference, DIMV = Direct Inference with Majority Voting}
\label{tab:main_result}
\setlength{\tabcolsep}{4pt}
\begin{tabular}{cccccc}
\hline
\toprule
\textbf{Model} & \makecell{\textbf{MMAU} \\ \textbf{test-mini}} & \makecell{\textbf{MMAU} \\ \textbf{test}} & \textbf{MMAR} & \textbf{MMSU} & \textbf{Avg.} \\ \midrule
\multicolumn{6}{@{}l}{\textbf{\emph{Qwen2.5-Omni 7B:}}} \\
DI & 72.40 & 70.60 & 57.70 & 56.84 & 64.39 \\ 
DIMV & 73.30 & 72.01 & 58.90 & 58.16 & 65.59 \\
SFT & 73.90 & 71.74 & 58.60 & 58.02 & 65.57 \\
Ours & \textbf{76.80} & \textbf{73.74} & \textbf{63.20} & \textbf{61.48} & \textbf{68.81} \\ \midrule
\multicolumn{6}{@{}l}{\textbf{\emph{Qwen2.5-Omni 3B:}}} \\
DI & 61.50 & 61.55 & 46.90 & 45.34 & 53.82 \\ 
DIMV & 63.90 & 62.59 & 46.70 & 47.90 & 55.27 \\ 
SFT & 65.20 & 62.84 & 48.20 & 48.74 & 56.25 \\
Ours & \textbf{72.30} & \textbf{71.05} & \textbf{57.90} & \textbf{58.18} & \textbf{64.86} \\ \bottomrule
\end{tabular}
\end{table}

We take Direct Inference (DI) of the unadapted model as a baseline and use reproduced results for a consistent inference setting.
As shown in Table~\ref{tab:main_result}, our method demonstrates its effectiveness by outperforming DI across all benchmarks with an average improvement of 4.42\% (7B) and 11.04\% (3B).  
Our method also outperforms DIMV (majority-vote inference, which provides pseudo-labels) and SFT trained on the same pseudo-labels.
We attribute this improvement to the ability of reinforcement learning to tolerate a certain degree of label inaccuracy and capture the underlying distribution rather than merely imitating.
Notably, the self-adapted 3B model surpasses the unadapted 7B model under direct inference on average (64.86 vs. 64.39).
These results demonstrate the potential of LALMs to self-improve under a label-free training paradigm at test time, and further suggest that small models can also achieve comparative performance without manual annotation.

\subsection{Ablation studies}
\label{ssec:ablation}

\vspace{-0.5em}
\begin{table}[ht]
\centering
\caption{Module ablation (7B): G-MV = GRPO with Majority Voting, C = Confidence-weighted advantage, MA = Multiple-attempt sampling}
\label{tab:parts_ablation}
\setlength{\tabcolsep}{4pt}
\begin{tabular}{lccccc}
\toprule
\textbf{Method} & \makecell{\textbf{MMAU} \\ \textbf{test-mini}} & \makecell{\textbf{MMAU} \\ \textbf{test}} & \textbf{MMAR} & \textbf{MMSU} & \textbf{Avg.} \\ \midrule
G-MV & 74.10 & 72.86 & 62.50 & 60.54 & 67.50 \\ \hline
+MA &  75.00 & \textbf{73.79} & 61.90 & 60.98 & 67.92 \\ \hline
+C & \textbf{76.80}& 72.45 & 62.90 & 60.40 & 68.14 \\ \hline
\makecell[l]{+C+MA \\ (Ours)} & \textbf{76.80} & 73.74 & \textbf{63.20} & \textbf{61.48} & \textbf{68.81} \\ \bottomrule
\end{tabular}
\end{table}

\textbf{Confidence-weighted advantage and multiple-attempt sampling.} 
Table~\ref{tab:parts_ablation} reports an ablation study of confidence-weighted advantage and multiple-attempt sampling.
GRPO with Majority Voting (G-MV) achieves the lowest average accuracy. 
Adding multiple-attempt sampling improves performance on most benchmarks, except MMAR.
Meanwhile, confidence weighting improves performance on MMAU test-mini and MMAR, but brings limited gains on larger datasets. 
When advantage collapse occurs, confidence weighting amplifies differences between samples and leads to more uneven updates.
The combination of these two modules consistently improves performance with an average gain of 1.31\%. 
Overall, the ablations indicate that confidence weighting and multiple-attempt sampling are complementary: the former reweights noisy pseudo-labels, while the latter reduces advantage-collapse updates, leading to the most consistent gains when combined.

\begin{table}[ht]
\centering
\caption{Audio-type-wise accuracy on MMAU test-mini with ablations on audio-type-restricted adaptation (7B).}
\label{tab:speech_type}
\setlength{\tabcolsep}{4pt}
\begin{tabular}{lccccc}
\toprule
\textbf{Method} & \textbf{Adaptation} & \textbf{Sound} & \textbf{Music} & \textbf{Speech} & \textbf{Avg.} \\ \midrule
DI & -- & 75.08 & 68.86 & 73.27 & 72.40 \\ \hline
DIMV & -- &77.48 & 69.16 & 73.27 & 73.30 \\ \hline
SFT & All & 78.38 & 69.16 & 74.17 & 73.90 \\ \hline
Ours & All & 82.88 & 72.16 & 75.38 & 76.80 \\ 
 & Sound & 80.78 & 69.76 & 75.98 & 75.50 \\ 
 & Music & 82.28 & 70.96 & 76.58 & 76.60 \\ 
 & Speech & 80.48 & 70.66 & 75.08 & 75.40 \\ \hline
\end{tabular}
\end{table}

\textbf{Audio-type-wise results and adaptation ablation.} We further analyze performance across different audio types and conduct ablations by restricting adaptation to each audio type, using a fixed number of update steps per subset for consistency.  Our method consistently outperforms DIMV and SFT across all audio types. Among different settings, music-only adaptation achieves the highest average accuracy, while single-type adaptation remains competitive with training-label baselines (DIMV). This suggests that reinforcement learning enables robust test-time adaptation without overfitting to a specific audio type.

\begin{figure}[t]
\includegraphics[width=\linewidth]{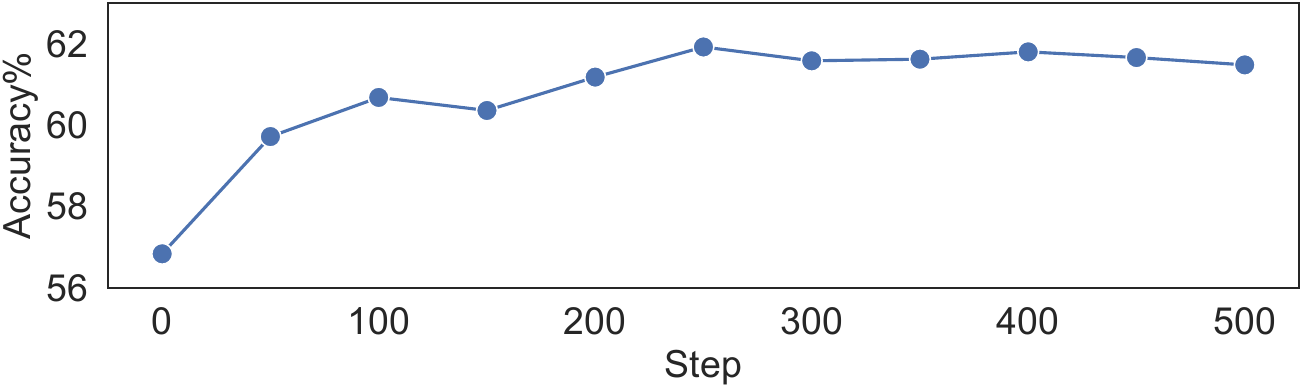}
\caption{Case study: accuracy over training steps (MMSU). 
We report accuracy at the fixed 500th update step for a consistent label-free protocol.  
Accuracy improves rapidly and largely stabilizes before a full pass over the test set (5,000 samples; global batch size 8), suggesting that adaptation on a subset of unlabeled test samples can already benefit the remaining samples from the same distribution.
}
\label{fig:future}
\end{figure}

\textbf{Effect of voting budget.} 
\begin{figure}[t]
\includegraphics[width=\linewidth]{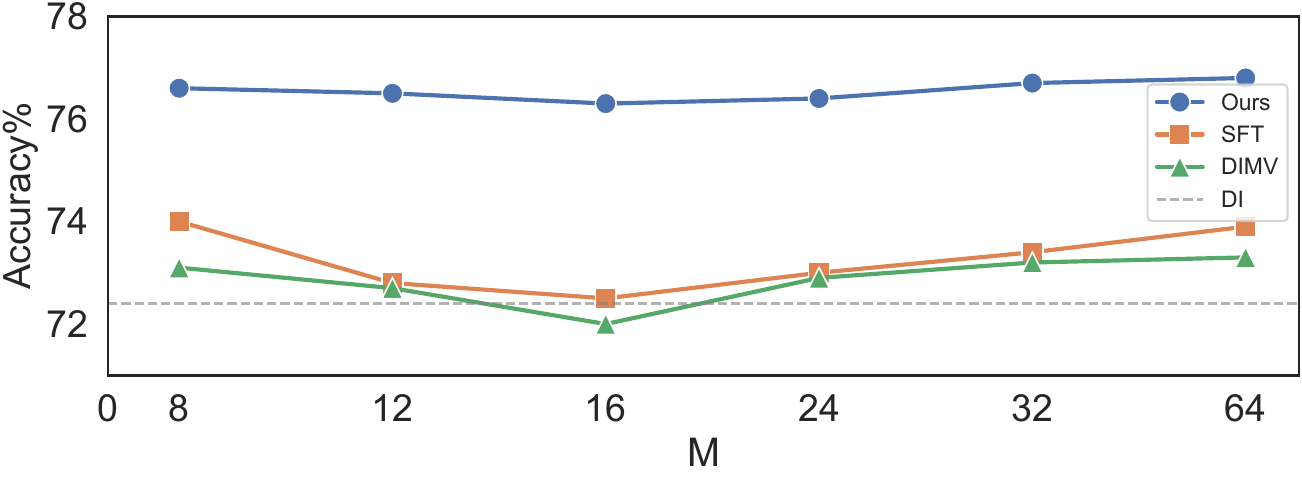}
\caption{Accuracy vs. $M$ on MMAU test-mini. $M$ is majority-vote sample count: inference-time voting for DIMV, adaptation-time pseudo-label voting for SFT/Ours (single-pass eval).}
\label{fig:votes}
\end{figure}
As shown in Fig.~\ref{fig:votes}, we sweep $M$, the number of samples used for majority voting when constructing pseudo-labels during adaptation.
For SFT/Ours, voting is applied only during adaptation, and inference after adaptation remains single-pass, whereas DIMV performs $M$-vote at inference.
Our method consistently outperforms both baselines across $M$.
While our method uses additional sampling during adaptation (e.g., GRPO rollouts),  it aims to convert multi-sample voting gains into improved single-pass performance.
Notably, with $M{=}8$ it reaches 76.6\% accuracy and surpasses DIMV even with $M{=}64$ inference-time votes, indicating that the gains do not rely on large $M$ at test time.
In contrast, DIMV is sensitive to $M$, requiring many votes to stabilize stochastic LLM outputs. 
Our method maintains strong performance even when the pseudo-label is unstable, owing to the RL framework and the confidence-weighted design. Beyond the effect of the voting budget, we further provide a case study of adaptation dynamics on MMSU in Fig.~\ref{fig:future}.

\subsection{Correlation between pseudo-label quality and confidence}
\label{ssec:analysis}

\begin{figure}[t]
\includegraphics[width=\linewidth]{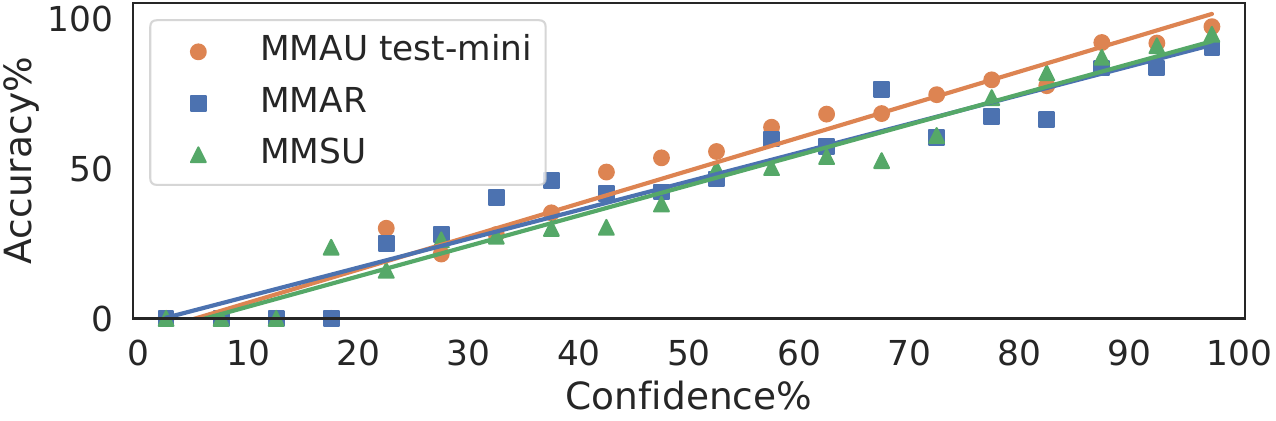}
\caption{Accuracy vs. Confidence of Pseudo-Label}
\label{fig:distribution}
\end{figure}
To examine the effectiveness of the confidence weighting mechanism, we investigate how pseudo-label accuracy varies with confidence.
We bin confidence into 5\% intervals, compute mean pseudo-label accuracy per bin, and fit a regression line.
Since the MMAU test labels are unavailable due to online evaluation, we report the remaining test datasets.
As shown in Fig.~\ref{fig:distribution}, a strong positive correlation between confidence and accuracy across different test datasets is observed, indicating that higher-confidence pseudo-labels are more reliable.
Since training signals obtained from higher-confidence pseudo-labels are more reliable, they should be encouraged to contribute more to the model updates.
The results indicate that the confidence weighting mechanism with reinforcement learning can effectively emphasize informative samples while tolerating noisy ones, aligning the model update with the inherent data distribution.

\section{Conclusion}
\label{sec:con}

We propose AQA-TTRL, a label-free test-time adaptation framework for audio question answering. 
By combining majority-vote pseudo-labeling with GRPO-based optimization, along with confidence weighting and multiple-attempt sampling, the framework enables stable self-improvement on unlabeled test data. 
Experiments on MMAU (test-mini/test), MMAR, and MMSU demonstrate consistent gains, with average improvements of 4.42\% on the Qwen2.5-Omni 7B and 11.04\% on 3B.
Notably, the self-improved 3B model surpasses the direct inference performance of the unadapted 7B model, indicating that targeted on-the-fly adaptation offers a parameter-efficient complement to model scaling.

\section{Generative AI Use Disclosure}
Generative AI tools were used solely for language polishing and clarity improvement.

\bibliographystyle{IEEEtran}
\bibliography{refs}

@inproceedings{gong2023listen,
  title={Listen, Think, and Understand},
  author={Gong, Yuan and Luo, Hongyin and Liu, Alexander H and Karlinsky, Leonid and Glass, James},
  booktitle={The Twelfth International Conference on Learning Representations (ICLR)},
  year={2024}
}

@inproceedings{gong_ltuas,
  title={Joint Audio and Speech Understanding},
  author={Gong, Yuan and Liu, Alexander H and Luo, Hongyin and Karlinsky, Leonid and Glass, James},
  year={2023},
  booktitle={2023 IEEE Automatic Speech Recognition and Understanding Workshop (ASRU)},
}

@inproceedings{
  tang2024salmonn,
  title={{SALMONN}: Towards Generic Hearing Abilities for Large Language Models},
  author={Changli Tang and Wenyi Yu and Guangzhi Sun and Xianzhao Chen and Tian Tan and Wei Li and Lu Lu and Zejun Ma and Chao Zhang},
  booktitle={The Twelfth International Conference on Learning Representations (ICLR)},
  year={2024},
}

@inproceedings{kong2024audio,
  title={{Audio Flamingo}: A Novel Audio Language Model with Few-Shot Learning and Dialogue Abilities},
  author={Kong, Zhifeng and Goel, Arushi and Badlani, Rohan and Ping, Wei and Valle, Rafael and Catanzaro, Bryan},
  booktitle={International Conference on Machine Learning (ICML)},
  year={2024},
}

@inproceedings{
  ghosh2025audio,
  title={{Audio Flamingo 2}: An Audio-Language Model with Long-Audio Understanding and Expert Reasoning Abilities},
  author={Ghosh, Sreyan and Kong, Zhifeng and Kumar, Sonal and Sakshi, S and Kim, Jaehyeon and Ping, Wei and Valle, Rafael and Manocha, Dinesh and Catanzaro, Bryan},
  booktitle={International Conference on Machine Learning (ICML)},
  year={2025},
}

@inproceedings{
  ghosh2025audio_3,
  title={{Audio Flamingo 3}: Advancing Audio Intelligence with Fully Open Large Audio Language Models},
  author={Sreyan Ghosh and Arushi Goel and Jaehyeon Kim and Sonal Kumar and Zhifeng Kong and Sang-gil Lee and Chao-Han Huck Yang and Ramani Duraiswami and Dinesh Manocha and Rafael Valle and Bryan Catanzaro},
  booktitle={Advances in Neural Information Processing Systems (NeurIPS)},
  year={2025},
}

@article{lu2025desta25Audio,
  title={{DeSTA2.5-Audio}: Toward General-Purpose Large Audio Language Model with Self-Generated Cross-Modal Alignment},
  author={Lu, Ke-Han and Chen, Zhehuai and Fu, Szu-Wei and Yang, Chao-Han Huck and Huang, Sung-Feng and Yang, Chih-Kai and Yu, Chee-En and Chen, Chun-Wei and Chen, Wei-Chih and Huang, Chien-yu and others},
  journal={arXiv preprint arXiv:2507.02768},
  year={2025}
}

@inproceedings{lu2025developing,
  title={Developing instruction-following speech language model without speech instruction-tuning data},
  author={Lu, Ke-Han and Chen, Zhehuai and Fu, Szu-Wei and Yang, Chao-Han Huck and Balam, Jagadeesh and Ginsburg, Boris and Wang, Yu-Chiang Frank and Lee, Hung-yi},
  booktitle={2025 IEEE International Conference on Acoustics, Speech and Signal Processing (ICASSP)},
  year={2025},
}

@inproceedings{lu24c_interspeech,
  title     = {{DeSTA}: Enhancing Speech Language Models through Descriptive Speech-Text Alignment},
  author    = {Ke-Han Lu and Zhehuai Chen and Szu-Wei Fu and He Huang and Boris Ginsburg and Yu-Chiang Frank Wang and Hung-yi Lee},
  year      = {2024},
  booktitle = {Interspeech},
  pages     = {4159--4163},
  doi       = {10.21437/Interspeech.2024-457},
  issn      = {2958-1796},
}

@article{Qwen-Audio,
  title={{Qwen-Audio}: Advancing Universal Audio Understanding via Unified Large-Scale Audio-Language Models},
  author={Chu, Yunfei and Xu, Jin and Zhou, Xiaohuan and Yang, Qian and Zhang, Shiliang and Yan, Zhijie  and Zhou, Chang and Zhou, Jingren},
  journal={arXiv preprint arXiv:2311.07919},
  year={2023}
}

@article{Qwen2-Audio,
  title={{Qwen2-Audio} Technical Report},
  author={Chu, Yunfei and Xu, Jin and Yang, Qian and Wei, Haojie and Wei, Xipin and Guo, Zhifang and Leng, Yichong and Lv, Yuanjun and He, Jinzheng and Lin, Junyang and others},
  journal={arXiv preprint arXiv:2407.10759},
  year={2024}
}

@article{Qwen2.5-Omni,
  title={{Qwen2.5-Omni} Technical Report},
  author={Xu, Jin and Guo, Zhifang and He, Jinzheng and Hu, Hangrui and He, Ting and Bai, Shuai and Chen, Keqin and Wang, Jialin and Fan, Yang and Dang, Kai and others},
  journal={arXiv preprint arXiv:2503.20215},
  year={2025}
}

@article{Qwen3-Omni,
  title={{Qwen3-Omni} Technical Report},
  author={Xu, Jin and Guo, Zhifang and Hu, Hangrui and Chu, Yunfei and Wang, Xiong and He, Jinzheng and Wang, Yuxuan and Shi, Xian and He, Ting and Zhu, Xinfa and others},
  journal={arXiv preprint arXiv:2509.17765},
  year={2025}
}

@article{huang2025step,
  title={{Step-Audio}: Unified understanding and generation in intelligent speech interaction},
  author={Huang, Ailin and Wu, Boyong and Wang, Bruce and Yan, Chao and Hu, Chen and Feng, Chengli and Tian, Fei and Shen, Feiyu and Li, Jingbei and Chen, Mingrui and others},
  journal={arXiv preprint arXiv:2502.11946},
  year={2025}
}

@article{wu2025step,
  title={{Step-Audio} 2 technical report},
  author={Wu, Boyong and Yan, Chao and Hu, Chen and Yi, Cheng and Feng, Chengli and Tian, Fei and Shen, Feiyu and Yu, Gang and Zhang, Haoyang and Li, Jingbei and others},
  journal={arXiv preprint arXiv:2507.16632},
  year={2025}
}

@article{deepseek-math,
  title={Deepseekmath: Pushing the limits of mathematical reasoning in open language models},
  author={Shao, Zhihong and Wang, Peiyi and Zhu, Qihao and Xu, Runxin and Song, Junxiao and Bi, Xiao and Zhang, Haowei and Zhang, Mingchuan and Li, YK and Wu, Yang and others},
  journal={arXiv preprint arXiv:2402.03300},
  year={2024}
}

@article{guo2025deepseek,
  title={Deepseek-r1: Incentivizing reasoning capability in {LLMs} via reinforcement learning},
  author={Guo, Daya and Yang, Dejian and Zhang, Haowei and Song, Junxiao and Zhang, Ruoyu and Xu, Runxin and Zhu, Qihao and Ma, Shirong and Wang, Peiyi and Bi, Xiao and others},
  journal={arXiv preprint arXiv:2501.12948},
  year={2025}
}

@article{chen2025seed,
  title={{SEED-GRPO}: Semantic Entropy Enhanced GRPO for Uncertainty-Aware Policy Optimization},
  author={Chen, Minghan and Chen, Guikun and Wang, Wenguan and Yang, Yi},
  journal={arXiv preprint arXiv:2505.12346},
  year={2025}
}

@inproceedings{rouditchenko2025omni,
  title={{Omni-R1}: Do You Really Need Audio to Fine-Tune Your Audio LLM?},
  author={Rouditchenko, Andrew and Bhati, Saurabhchand and Araujo, Edson and Thomas, Samuel and Kuehne, Hilde and Feris, Rogerio and Glass, James},
  booktitle = {2025 IEEE Automatic Speech Recognition and Understanding Workshop (ASRU)},
  year      = {2025},
}

@article{shao2025spurious,
  title={Spurious Rewards: Rethinking Training Signals in {RLVR}},
  author={Shao, Rulin and Li, Shuyue Stella and Xin, Rui and Geng, Scott and Wang, Yiping and Oh, Sewoong and Du, Simon Shaolei and Lambert, Nathan and Min, Sewon and Krishna, Ranjay and others},
  journal={arXiv preprint arXiv:2506.10947},
  year={2025}
}

@article{li2025reinforcement,
  title={Reinforcement learning outperforms supervised fine-tuning: A case study on audio question answering},
  author={Li, Gang and Liu, Jizhong and Dinkel, Heinrich and Niu, Yadong and Zhang, Junbo and Luan, Jian},
  journal={arXiv preprint arXiv:2503.11197},
  year={2025}
}

@inproceedings{zuo2025ttrl,
  title={{TTRL}: Test-Time Reinforcement Learning},
author={Zuo, Yuxin and Zhang, Kaiyan and Sheng, Li and Qu, Shang and Cui, Ganqu and Zhu, Xuekai and Li, Haozhan and Zhang, Yuchen and Long, Xinwei and Hua, Ermo and others},
  booktitle={Advances in Neural Information Processing Systems (NeurIPS)},
  year={2025}
}

@inproceedings{
sakshi2025mmau,
title={{MMAU}: A Massive Multi-Task Audio Understanding and Reasoning Benchmark},
author={S Sakshi and Utkarsh Tyagi and Sonal Kumar and Ashish Seth and Ramaneswaran Selvakumar and Oriol Nieto and Ramani Duraiswami and Sreyan Ghosh and Dinesh Manocha},
booktitle={The Thirteenth International Conference on Learning Representations (ICLR)},
year={2025},
}

@inproceedings{ma2025mmar,
  title={{MMAR}: A Challenging Benchmark for Deep Reasoning in Speech, Audio, Music, and Their Mix},
  author={Ma, Ziyang and Ma, Yinghao and Zhu, Yanqiao and Yang, Chen and Chao, Yi-Wen and Xu, Ruiyang and others},
  booktitle={Advances in Neural Information Processing Systems (NeurIPS)},
  year={2025}
}

@inproceedings{
    wang2025mmsu,
    title={{MMSU}: A Massive Multi-task Spoken Language Understanding and Reasoning Benchmark},
    author={Dingdong WANG and Jincenzi Wu and Junan Li and Dongchao Yang and Xueyuan Chen and Tianhua Zhang and Helen M. Meng},
    booktitle={The Fourteenth International Conference on Learning Representations (ICLR)},
    year={2026},
}

@inproceedings{wang2022self,
  title={Self-consistency improves chain of thought reasoning in language models},
  author={Wang, Xuezhi and Wei, Jason and Schuurmans, Dale and Le, Quoc and Chi, Ed and Narang, Sharan and Chowdhery, Aakanksha and Zhou, Denny},
  booktitle={The Eleventh International Conference on Learning Representations (ICLR)},
  year={2023}
}

@inproceedings{
zhang2026scafgrpo,
title={Scaf-{GRPO}: Scaffolded Group Relative Policy Optimization for Enhancing {LLM} Reasoning},
author={Xichen Zhang and Sitong Wu and Yinghao Zhu and Haoru Tan and Shaozuo Yu and Ziyi He and Jiaya Jia},
booktitle={The Fourteenth International Conference on Learning Representations (ICLR)},
year={2026},
}

\end{document}